\documentclass[mathleft,fleqn,
]{an}
\usepackage[compatibility=false]{caption}
\usepackage{subcaption}
\usepackage{graphicx}
\usepackage[varg]{txfonts}
\usepackage{natbib}
\bibpunct{(}{)}{;}{a}{}{,}
\begin{document}

\Yearpublication{2018}%
\Yearsubmission{2017}%

\title{Photometry of asteroids (5141), (43032), (85953), (259221) and (363599) observed at Pic du Midi Observatory
        }

\author{B.\,A.\, Dumitru\inst{1,2,3,4}
\and  M.\, Birlan\inst{1,3}
\and  A.\, Sonka\inst{3,4}
\and  F.\, Colas\inst{1}
\and  D. \,A.\, Nedelcu\inst{3,1}
}
\titlerunning{Photometry of asteroids observed at Pic du Midi Observatory}

\institute{
Institut de M\'ecanique C\'eleste et des Calculs des \'Eph\'em\'erides, CNRS UMR8028, Paris Observatory, PSL Research University, 77 av Denfert Rochereau,
75014 Paris cedex, France\\
				\email{Mirel.Birlan@obspm.fr}
\and
Institute of Space Science (ISS), 409 Atomistilor str, 077125 Magurele, Ilfov, Romania\\
             	\email{bogdan.dumitru@spacescience.ro}
\and
Astronomical Institute of Romanian Academy, 5-Cu\c titul de Argint, 040557 Bucharest, Romania
\and
Faculty of Physics, Bucharest University, 405 Atomistilor str, 077125 Magurele, Ilfov, Romania
}

\received{19 December 2017}
\accepted{7 March 2018}
\publonline{XXXX}


\abstract{%
	These are the first results of an observational program devoted to complete physical data of asteroids that could produce or feed meteoroid streams. Our results are based on the optical observation at Pic du Midi observatory in  April  6-7, 2016 and January 17-18, 2018. We will present the lightcurve of asteroid (259221)~2003~BA21 associated Daytime Sextantids (221 DSX) and November $\theta$ Aurigids (390 THA), the lightcurves and colors of asteroid (363599)~2004~FG11, associated with Daytime $\zeta$ Perseids (172 ZPE) as well as $g-r$ and $g-i$ and the corresponding reflectances for the (85953)~1999~FK21 associated with Daytime $\xi$ Sagittariids (100 XSA). Also we will present the colors for two additional objects, namely (5141)~Tachibana and (43032)~1999~VR26.
}
 
\maketitle

\section{Introduction}

The study of the links between comets, asteroids, and meteor streams can help the understanding of several scientific aspects 
such as the formation and evolution of our Solar System \citep{1973ApJ...183.1051G} and the appearance of life on Earth.
Indeed, two of the most important constituents of life, water and carbon, are present on over 60\% of asteroids \citep{1996Icar..124..352B, 1987Icar...72..304B}. 
Giving the lessons of Tunguska \citep{2001A&A...377.1081F, 1938ASPL....3...78K, 1930fcsm.book.....S} and Chelyabinsk events \citep{2013Sci...342.1069P}, these studies
are prerequisites for any of the current mitigation methods (gravitational tractor, kinetic impactor of explosive deflection).
Space mining is another aspect which can be interesting for human civilization  in the future efforts of Solar System colonization \citep{1996msur.book.....L}.

Nowadays  a link between comets, asteroids and meteor streams is firmly established \citep{2013MNRAS.430.2377J}. After the confirmation of Weiss prediction from 1963 about a large storm of Andromedids meteor stream associated with comet 3D/Biela \citep{2011A&G....52b..20W}, the discovery of asteroids as extinct comets \citep{1989aste.conf..880W} and asteroids with activity \citep{2015aste.book..221J}, thus blurring the distinction between the two classes, this concept was largely accepted.

A systematic research of asteroids that could produce meteors based on both dynamical and physical similarities between asteroids, meteor showers and meteorites was recently started \citep{2017A&A...607A...5D}. The authors produced a list of objects, mainly Near Earth Asteroids (NEAs),  which can be associated with meteor showers. The physical data on candidate asteroids is however scarce.

The aim of this paper is to identify, based on their physical  properties, the asteroids that can produce or feed the meteor showers.

We developed an observational project dedicated to the acquisition of relevant physical data on the candidate asteroids (lightcurves, colors and spectra). This data will allow a solid statistical analysis of the aforementioned links while enlarging the sample of NEA with known physical parameters.

The lightcurve is  the primary way to determine rotational properties of an asteroid. From the rotation period we can determine if an object has a ruble-pile or  monolithic structure \citep{2006Icar..181...63P}. From several lightcurves obtained at several oppositions, the absolute magnitude, the shape, and pole of the object could be obtained.

The asteroids colors are used to determine some characteristics of asteroid's surface and for a first order estimation of its taxonomic type \citep{2000Icar..146..204F}. The systems of filters, commonly used are Johnson-Cousins U, B, V, R and I \citep[see][]{1979PASP...91..589B, 1974MNRAS.166..711C, 1953ApJ...117..313J} and Sloan Digital Sky Survey (SDSS) u, g, r, i and z \citep{2000AJ....120.1579Y}.

Our sample of asteroids (and NEAs in general) is generally difficult  to observe. The observing window of opportunity for a NEA is approximately one or two weeks during their close approach to Earth. On average, this favorable geometry occurs five times per century \citep{2015A&A...581A...3B}.

\section{Observations}

	Our observations were obtained at Pic du Midi observatory from Pyrenees mountains, France located at 2\,870m of altitude. The observations were made in April 6-7, 2016 and January 17-18, 2018 using the T1M 1.05~m telescope, an iKon-L Andor CCD camera with a 2k X 2k E2V chip (pixel scale 0.22 "/pix) and SDSS filters \citep{2013P&SS...85..299V}. We used the 2x2 binning mode in order to avoid the oversampling of images. The seeing was not constant during our run with FWHM between 1.2 and 2 arcsec.

Our main targets where (363599)~2004~FG11, (85953)~1999~FK21 and (259221)~2003~BA21, previously associated with Daytime $\zeta$ Perseids (172 ZPE), Daytime $\xi$ Sagittariids (100 XSA) and, respectively with two meteor showers, Daytime Sextantids (221 DSX) and November $\theta$ Aurigids (390 THA) \citep{2017A&A...607A...5D}. The backup target of our run where the main belt asteroids (5141)~Tachibana and (43032)~1999~VR26.

\section{Analysis Methods}

	Reference stars magnitude were obtained from Aladin\footnote{http://aladin.u-strasbg.fr/} sky atlas \citep{2014ASPC..485..277B, 2000A&AS..143...33B} with SDSS Photometry Catalog released in 2012.

	Aladin allows the user to visualize digitized astronomical images or full surveys, superimpose entries from astronomical catalogs or databases, and interactively access related data and information from the Simbad database \citep{2000A&AS..143....9W}, the VizieR service \citep{2000A&AS..143...23O} and other archives for all known astronomical objects in the field.

\subsection{Colors and reflectance extraction} \label{sec:color}

	The filter used in the run where SDSS $u$ = 0.354 nm, $g$ = 0.477 nm, $r$ = 0.623 nm, $i$ = 0.763 nm and $z$ = 0.913. The targets were not detected in  $u$ and $z$ filters. 
	
	For each asteroid were computed the reflectance colors $g-r$, $g-i$ and $\log{reflectance}$. The computation method was taken from EAR-A-I0035-5-SDSSTAX-V1.1 database

	The reflectance color (C) is:
{\setlength{\mathindent}{0pt}
\begin{equation}
  \label{diference}
  C = (M_{1} - M_{2})  - C_{S}
\end{equation}}%
	where $M_1$ and $M_2$ are the magnitudes of the object in the two filters  and $C_{S}$ is the color of the Sun ($g-r$ = 0.45 $\pm$ 0.02 and $g-i$ = 0.55 $\pm$ 0.03) obtained from \citet{2001AJ....122.2749I}.

	The reflectance color error ($\delta_{Color}$) is:
{\setlength{\mathindent}{0pt}
\begin{equation}
  \label{error}
  \delta_{C} = \sqrt{\delta_{M_{1}}^2 + \delta_{M_{2}}^2 + \delta_{C_{S}}^2}
\end{equation}}%
	where $\delta_{M_{1} }$ and $\delta_{M_{2} }$ are the standard deviations for each filter and $\delta_{C_{S} }$ is the standard deviations of solar color.

	Reflectance $R_{C}$ is:
{\setlength{\mathindent}{0pt}
\begin{equation}
  \label{reflectance}
  \log{R_{C}} = 0.4*C + \log(Rref)
\end{equation}}%
derived from Pogson equation, used in \citet{2010A&A...510A..43C} for computing the $\log{reflectances}$. We use this $\log{reflectance}$ because we will compare our results with the data from database 
to associate our objects to a taxonomic class.

The error $\delta_{R_{C}}$ of $R_{C}$  was obtained by:
{\setlength{\mathindent}{0pt}
\begin{equation}
  \label{reflectance_error}
  \delta_{R_{C}} = 0.4* \delta_{C}
\end{equation}}%

\subsection{Lightcurve} 

	The composite lightcurves were obtained using the following procedure:
	 
	(1) The apparent magnitude $M_{a}$ of the asteroid was obtained  using  three reference stars by:
{\setlength{\mathindent}{0pt}
\begin{equation}
  \label{app_mag}
  M_{a} = \left(M_{ia} - \frac{\sum M_{is}}{3}\right)  + \frac{\sum M_{rs}}{3}
\end{equation}}%

where $M_{ia}$ and $M_{is}$ are the instrumental magnitude of the asteroid and reference stars, and $M_{rs}$ is the magnitude of the reference stats from SDSS catalog.

	(2) Reduced magnitude $M_{R}$ of the asteroid,  magnitude at distance of 1 a.u. from Sun and Earth, is:
{\setlength{\mathindent}{0pt}
\begin{equation}
  \label{red_mag}
  M_{R} = M_{a} - 5*\log(r*d)
\end{equation}}%

where $r$ and $d$ are the heliocentric and geocentric distances of the asteroid.

	(3) To obtain the composite lightcurve we need to remove the influence of the phase angle from the reduced magnitude (i.e. the absolute magnitude). Considering a linear relationship between the reduced magnitude and the phase angle (Fig.~\ref{fig:abs}), we  obtain the absolute magnitude:
{\setlength{\mathindent}{0pt}
\begin{equation}
  \label{abs_mag}
  M_{SC} = M_{a} - l * PA
\end{equation}}%

where $l$ is the linear plot slope and $PA$ is the phase angle.

	(4) Rotational period and composite lightcurve.
	To obtain the rotation period we use a method for non-equidistant time series data, namely the Lomb-Scargle periodogram \citep{1982ApJ...263..835S}. This method estimates the period with a sinusoidal fit function (see Fig.~\ref{fig:LS}). Finally, all data is normalized to a single rotation period.

\section{Results} \label{sec:rez}

	We obtained colors for all asteroids and two lightcurves of NEA (363599)~2004~FG11 and (259221)~2003~BA21. The data for the lightcurves were obtained, for (363599)~2004~FG11, during two nights in April 6-7, 2016 for a total observing time of approximately 4 h (1.6 h during the first night and 2.5 h during the second night) in SDSS $r$ filter and for (259221)~2003~BA21 during the night between January 17-18, 2018 for a total observation time of approximately 6 h in SDSS $g$ filter. The colors for all objects were obtained in  April 6, 2016.

\subsection{What we know} \label{sec:ref}

	We investigated the literature for previous physical studied of our targets. The search was done in databases from Asteroid Lightcurve Photometry (ALCDEF\footnote{http://alcdef.org/}), Near Earth Objects Dynamic Site (NEODyS\footnote{http://newton.dm.unipi.it/neodys/}), Small Bodies Data Ferret\footnote{http://sbntools.psi.edu/ferret/}, Small Main-Belt Asteroid Spectroscopic Survey (SMASS\footnote{http://smass.mit.edu/catalog.php}) and European Asteroids Research Node (EARN\footnote{http://earn.dlr.de/nea/}).

	Our findings are presented in Table~\ref{literature_tab}.
	
\begin{table*}
\caption{Available physical data of our targets}             
\label{literature_tab}      
\centering         
\begin{tabular}{c c c c c c c c} 
\hline\hline                           
Number & Name & Orbit type & Diameter[km] & Rotational period [h] & Albedo & Taxonomy & Reference \\  \hline
5141  & Tachibana & Main belt & 8.709 $\pm$ 0.257 & --                   & 0.212 $\pm$ 0.043 & --     & 1, -- , 1, -- \\
43032 & 1999 VR26 & Main belt & 4.897 $\pm$ 0.887 & 32.890 $\pm$ 0.078   & 0.155 $\pm$ 0.082 & --     & 2, 3, 2,-- \\
85953 & 1999 FK21 & Athen     & 0.59              & 17.62 $\pm$ 0.05     & 0.32              & S-type & 4, 5, 4, 6 \\
363599& 2004 FG11 & Appolo    & 0.152 $\pm$ 0.003 & $<$ 4 , 22 $\pm$ 0.5 & 0.306 $\pm$ 0.050 & V-type & 7, 8 and 9, 7, 10 and 11 \\
259221 & 2003 BA21 & Appolo & 0.435 - 0.973 & -- & -- & -- & -- \\
\hline
\end{tabular}

 \begin{minipage}{\linewidth}
    \ignorespaces
(1)~\citet{2011ApJ...741...68M}; 
(2)~\citet{2015A&A...578A..42R}; 
(3)~\citet{2016Icar..269...15Y}; 
(4)~\citet{2003Icar..166..116D};  
(5)~\citet{2016MPBu...43..240W}; 
(6)~\citet{2004Icar..170..259B}; 
(7)~\citet{2014ApJ...784..110M}; 
(8)~\citet{2012CBET.3091....1T}; 		 
(9)~\citet{2014MPBu...41..213W};
(10)~\citet{2010ATel.2571....1H};
(11)~\citet{2010DPS....42.1316S};
  \end{minipage}%
\end{table*}

     Two of the asteroids may be in a tumbling spin state - (85953) 1999 FK21 \citep{2016MPBu...43..240W} and (43032) 1999 VR26 \citep{2016Icar..269...15Y}. (363599) 2004 FG11 is a confirmed binary system. (259221)~2003~BA21 is an Apollo asteroid with large eccentricity. {\bf Its perihelion is interior to Mercury orbit. }

\subsection{(363599) 2004 FG11 lightcurve}

\begin{figure}
\includegraphics[width=\linewidth]{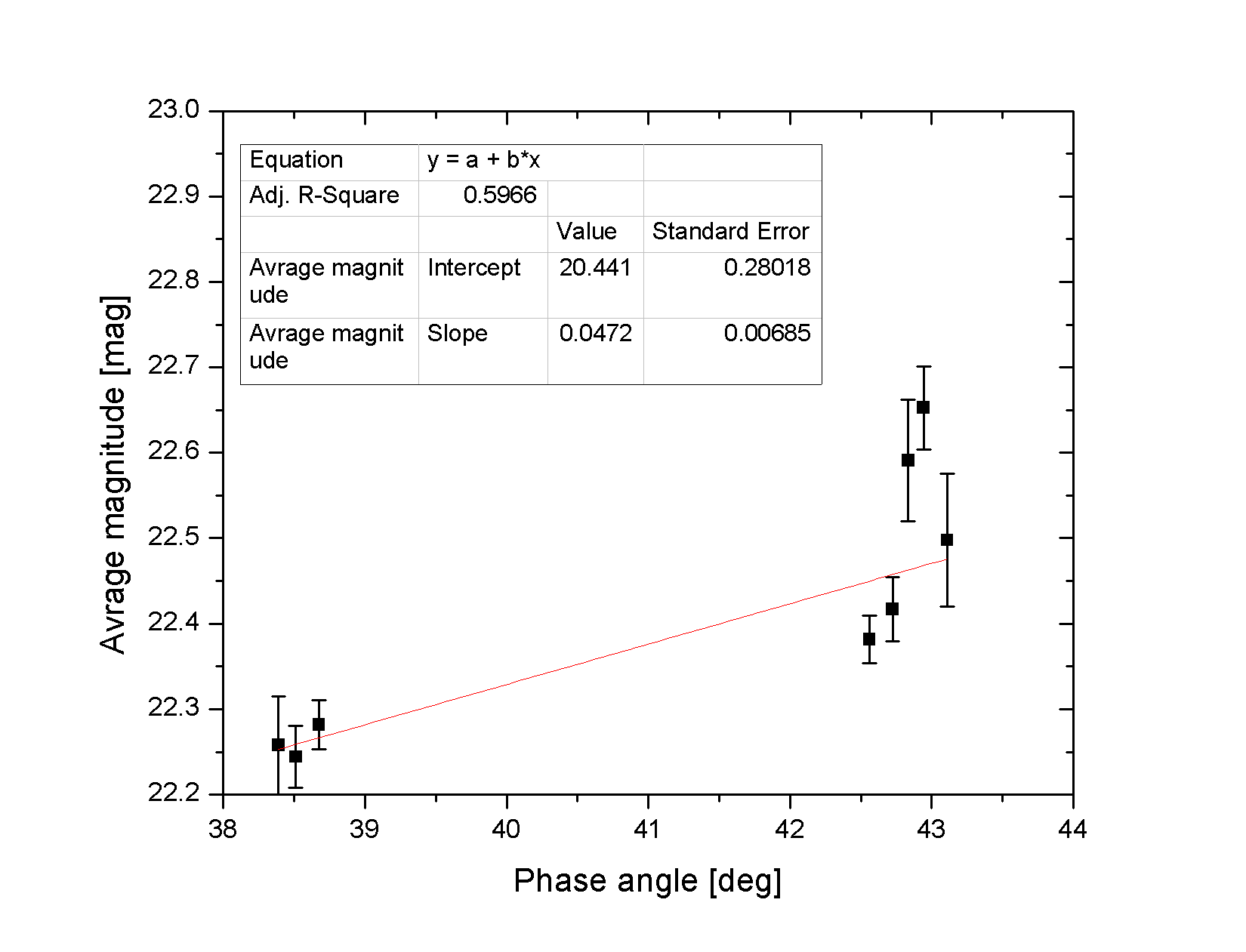}
\caption{ The representation of reduced magnitude versus phase angle.}
\label{fig:abs}
\end{figure}

	In the first step, we compute the absolute magnitude of the asteroid and we extract the phase angle dependence from our data using Eq.~\ref{abs_mag} (Fig.~\ref{fig:abs})

\begin{figure}
\includegraphics[width=\linewidth]{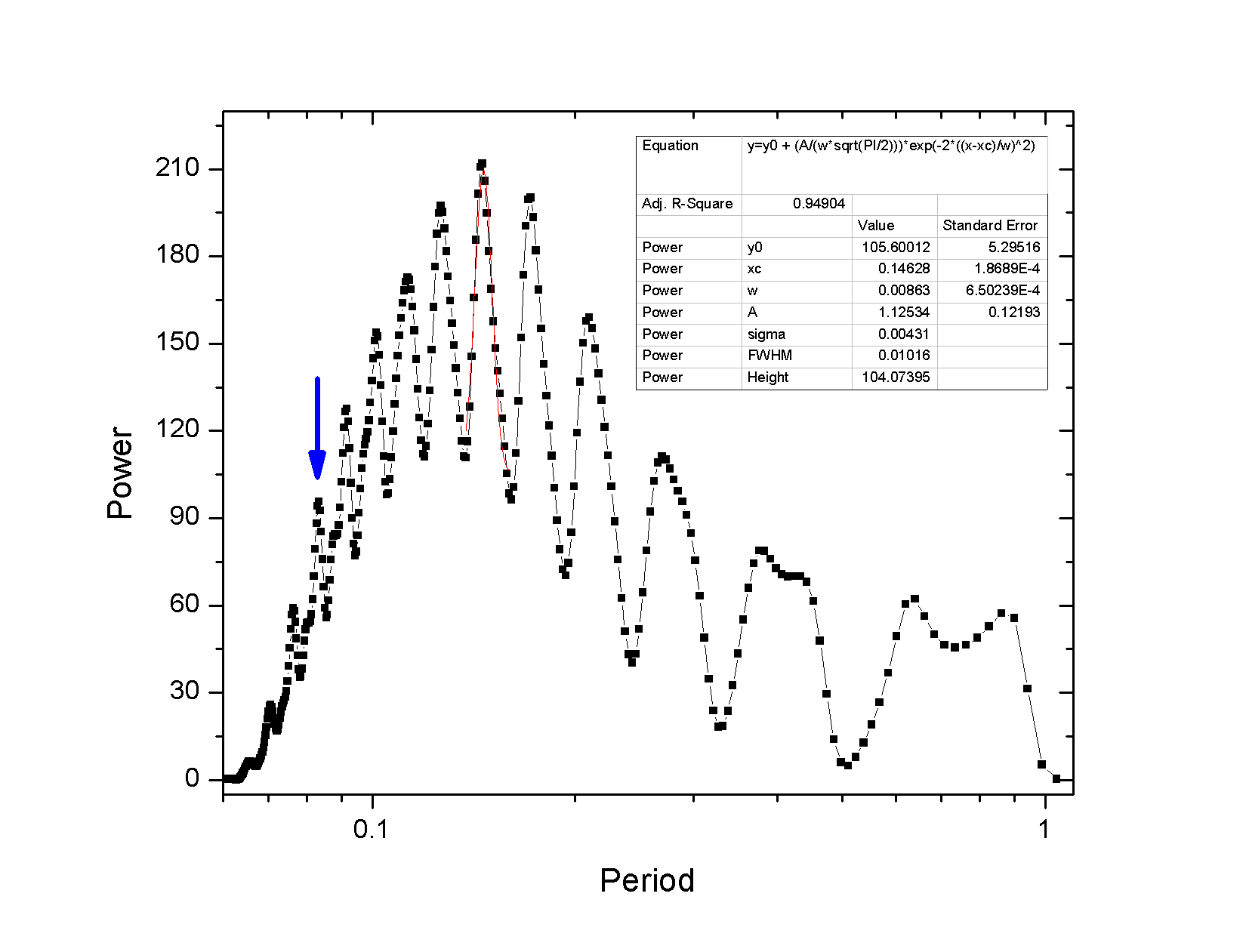}
\caption{Periodogram for (363599)~2004~FG11. The arrow represents the closest peak to the 4~h rotational period of primary (available from literature).}
\label{fig:LS}
\end{figure}

	We obtained an absolute magnitude of 20.441 $\pm$ 0.28 in r filter ( corresponding to a V filter transformation 20.7 $\pm$ 0.4).
	For the Lomb-Scargle analysis we used the NASA Exoplanet Archive Periodogram Service\footnote{https://exoplanetarchive.ipac.caltech.edu} (see Fig.~\ref{fig:LS}).

	The most probable period inferred from our observations of (363599) 2004 FG11 is of 0.2926 $\pm$ 0.0004 days (7.021 $\pm$ 0.001 h). This solution does not agree with the one from the literature where the estimation of rotational period for the primary is shorter the 4 h and the period of the binary system is  22 $\pm$ 0.5 h.

\begin{figure*}
\centering

\begin{subfigure}{.48\textwidth}
	\centering
\includegraphics[width=\hsize]{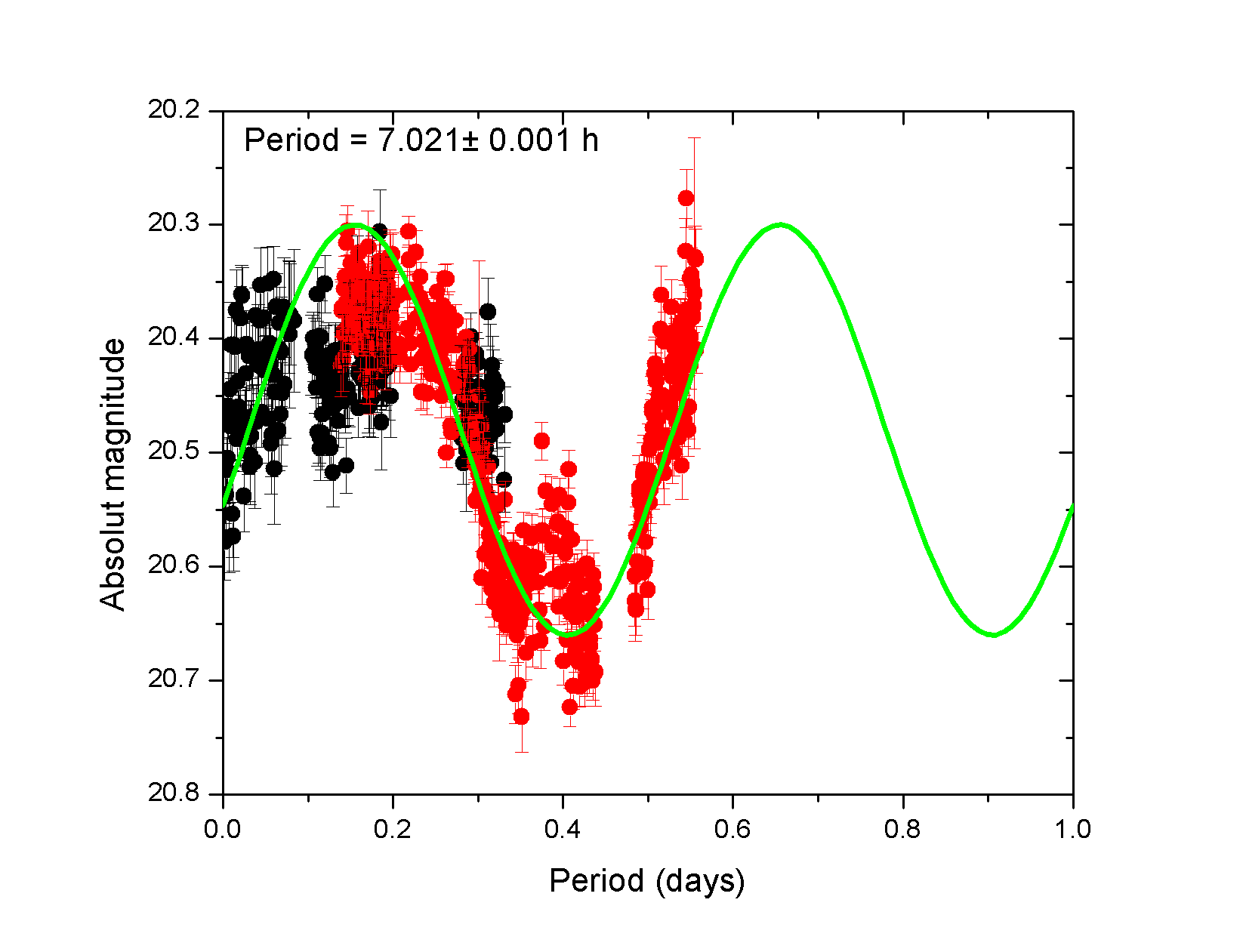}
\caption{}
\label{lightcurve1}
\end{subfigure}
\begin{subfigure}{.48\textwidth}
	\centering
\includegraphics[width=\hsize]{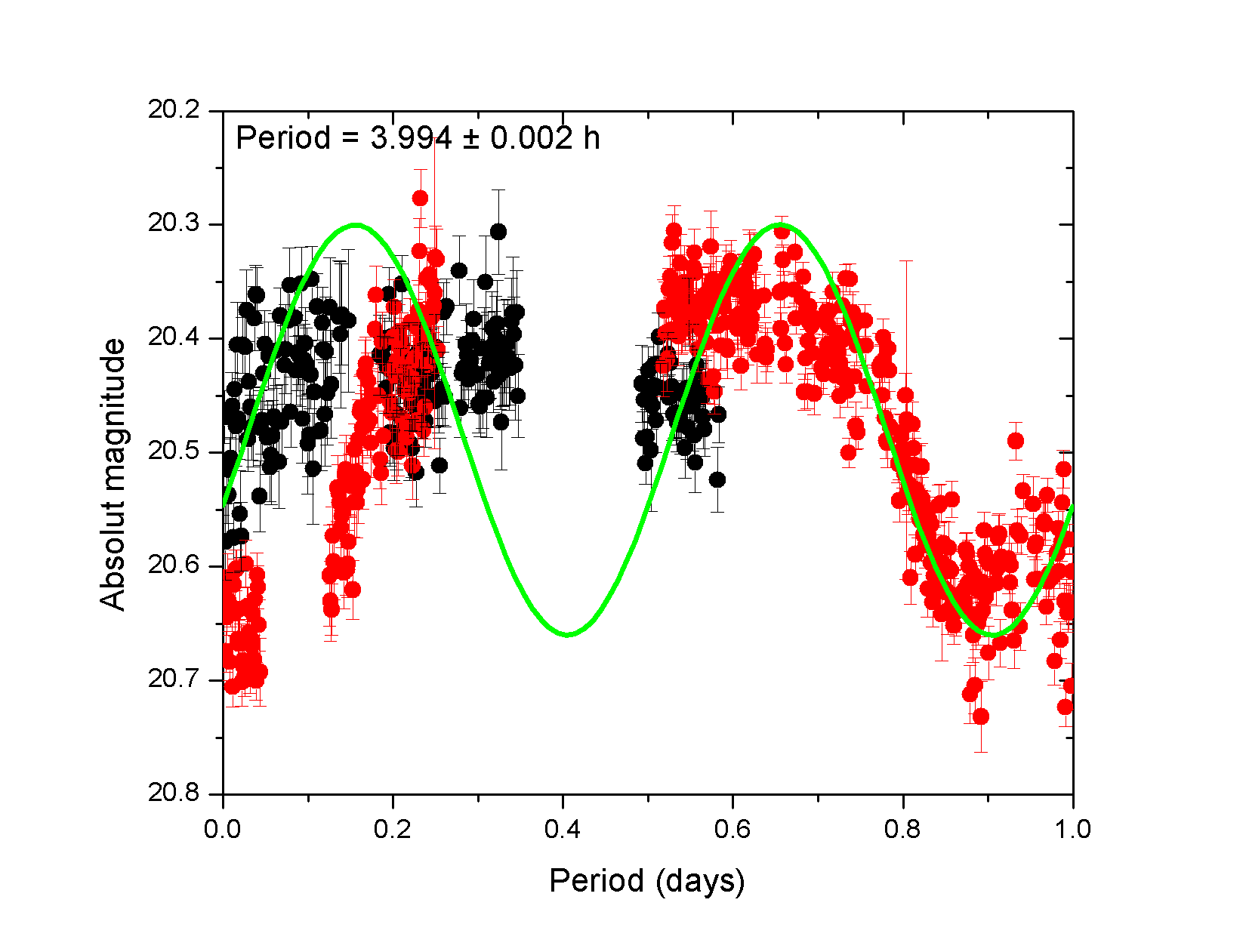}
\caption{}
\label{lightcurve2}
\end{subfigure}

	\caption{(363599)~2004~FG11 composite lightcurve. The black and red colors represent the nights when data were taken (black - 2016 April 06 and red - 2016 April 07). In Fig.~\ref{lightcurve1} is represented the lightcurve with a period of 7.021 $\pm$ 0.001 h. In Fig.~\ref{lightcurve2} is represented the lightcurve with a period of 3.994 $\pm$ 0.002 h}
		\label{lightcurve}
\end{figure*}

	As a check, we pack the lightcurve data using the period inferred from the peak closest to a 4h period (the arrow in Fig.~\ref{fig:LS}). These two lightcurves are presented in Fig.~\ref{lightcurve}. While the visual inspection  favors the 7 hours period, the period of 4 hours could not be completely excluded.

\begin{figure}
\centering
\includegraphics[width=\hsize]{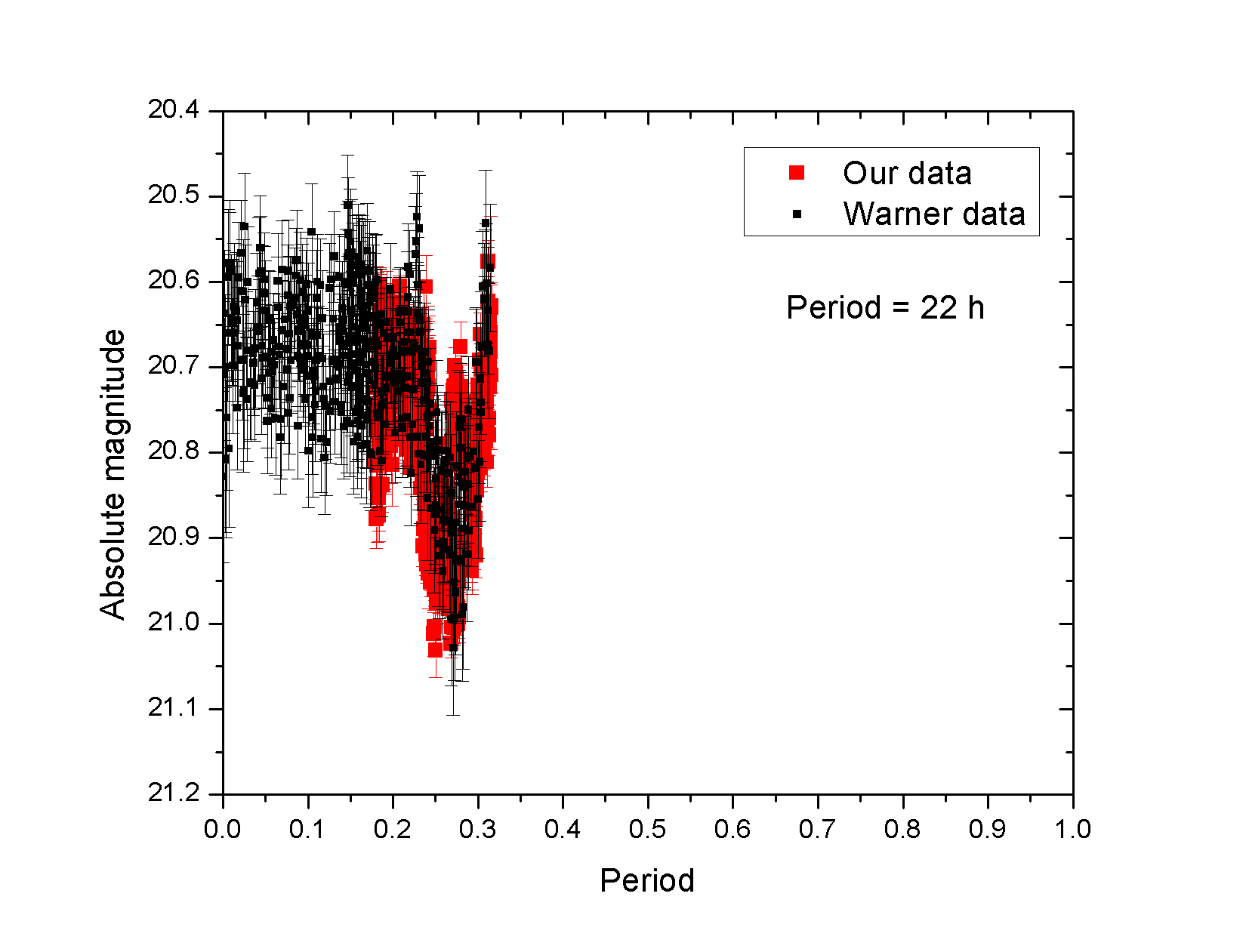}
	\caption{A comparison between our data and B. Warner data.}
		\label{test}
\end{figure}

	The period of the binary system was investigated using the data previously published by \cite{2014MPBu...41..213W}. For this, we manually set the period of the binary system to 22 h and we made a direct comparison of the results. The results are presented in Fig.~\ref{test}. Our data is overlapping the same region of the rotational period as the one from the literature.

	The result of our comparison is presented in Fig.~\ref{test} and we notice the good agreement between our data and those of the literature.

	From this test alone we can not draw a definitive conclusion. Our data agrees well with the known period of the binary system, but we do not find the 4~h period of the primary. 
	More observational data is needed for refining the rotational parameters of this binary system.

\subsection{(259221) 2003 BA21 lightcurve}

\begin{figure}
\includegraphics[width=\linewidth]{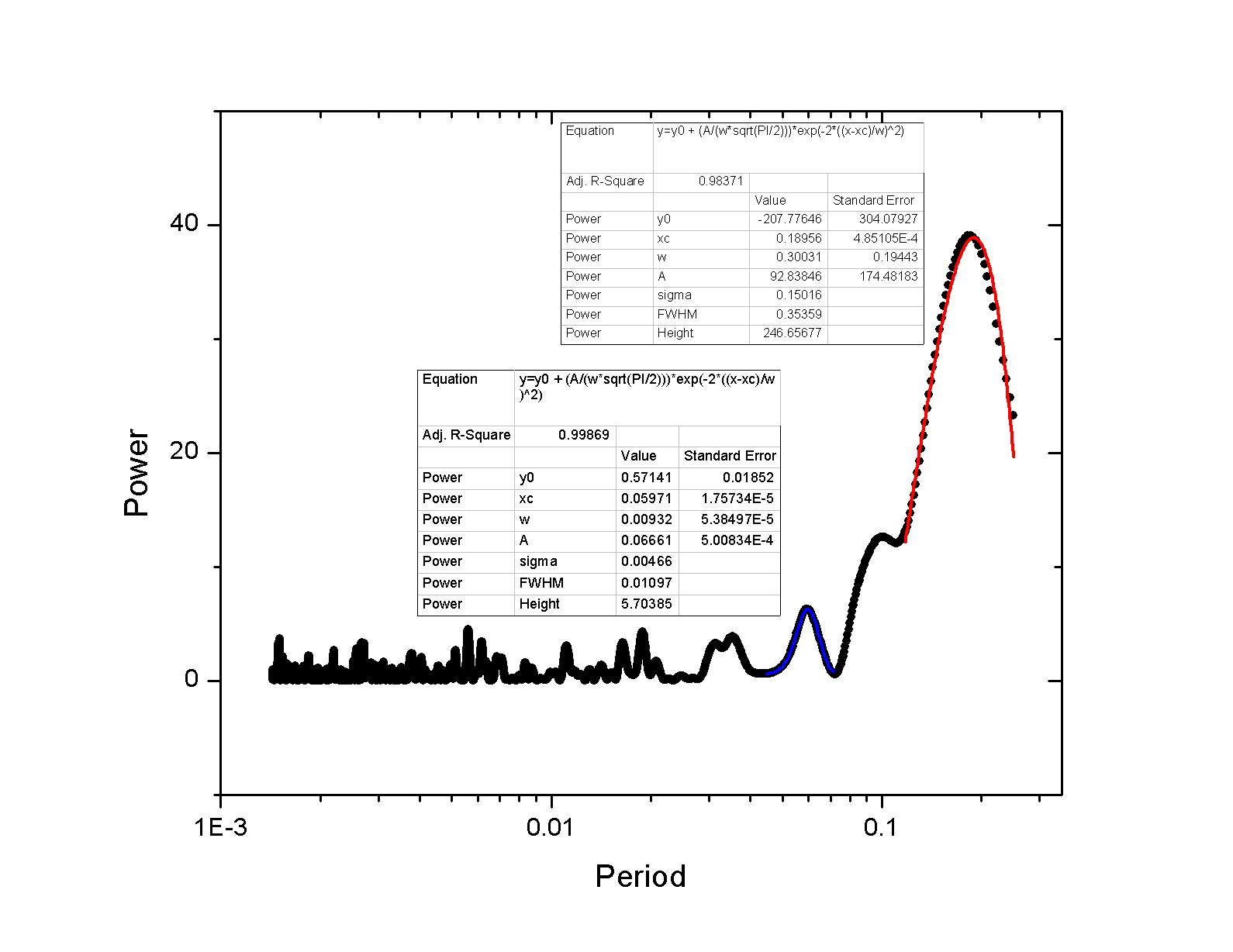}
\caption{Periodogram for (259221) 2003 BA21.}
\label{fig:LS_259221}
\end{figure}

	Fig.~\ref{fig:LS_259221} present the periodogram of this asteroid.

\begin{figure*}
\centering

\begin{subfigure}{.48\textwidth}
	\centering
\includegraphics[width=\hsize]{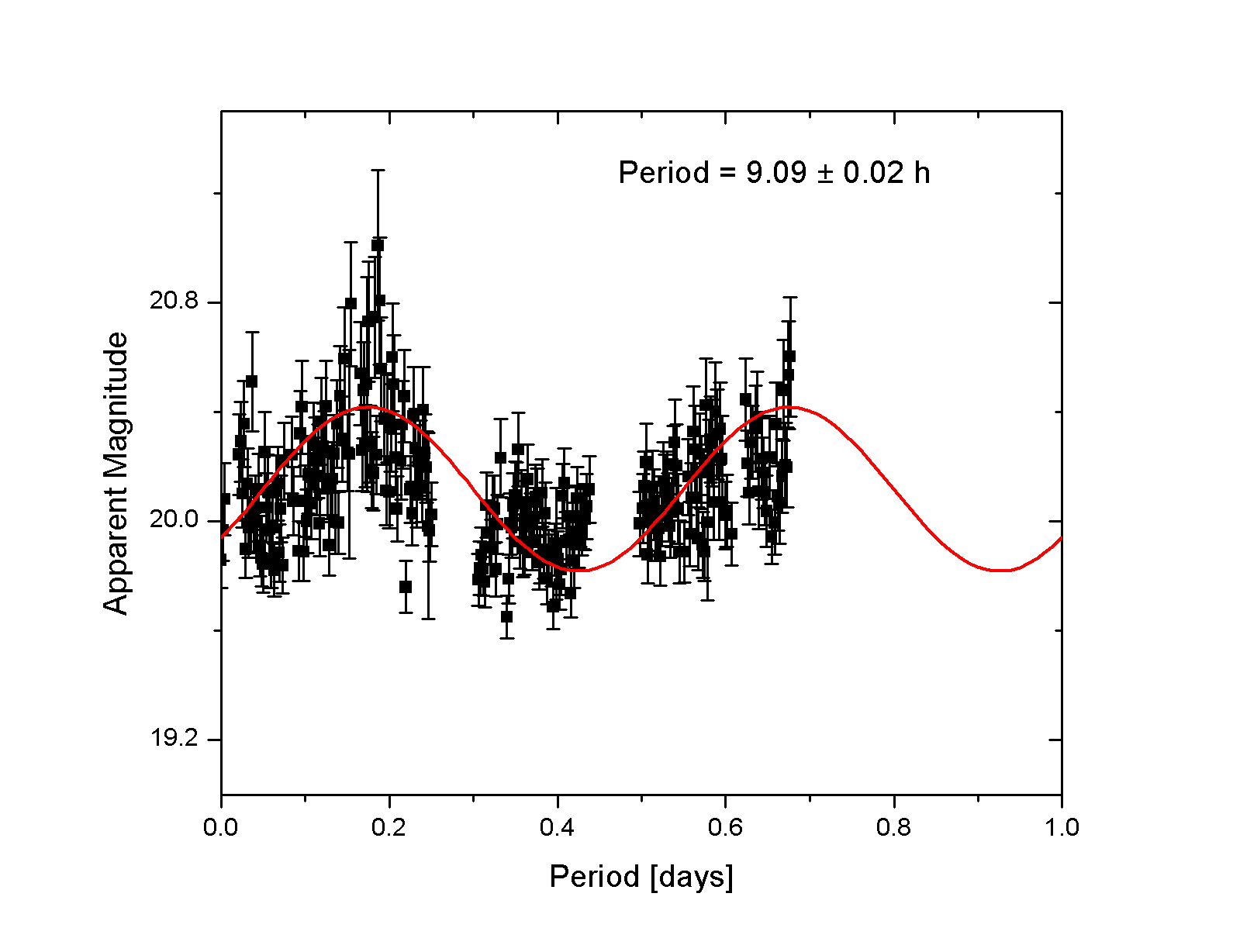}
\caption{}
\label{lightcurve2.1}
\end{subfigure}
\begin{subfigure}{.48\textwidth}
	\centering
\includegraphics[width=\hsize]{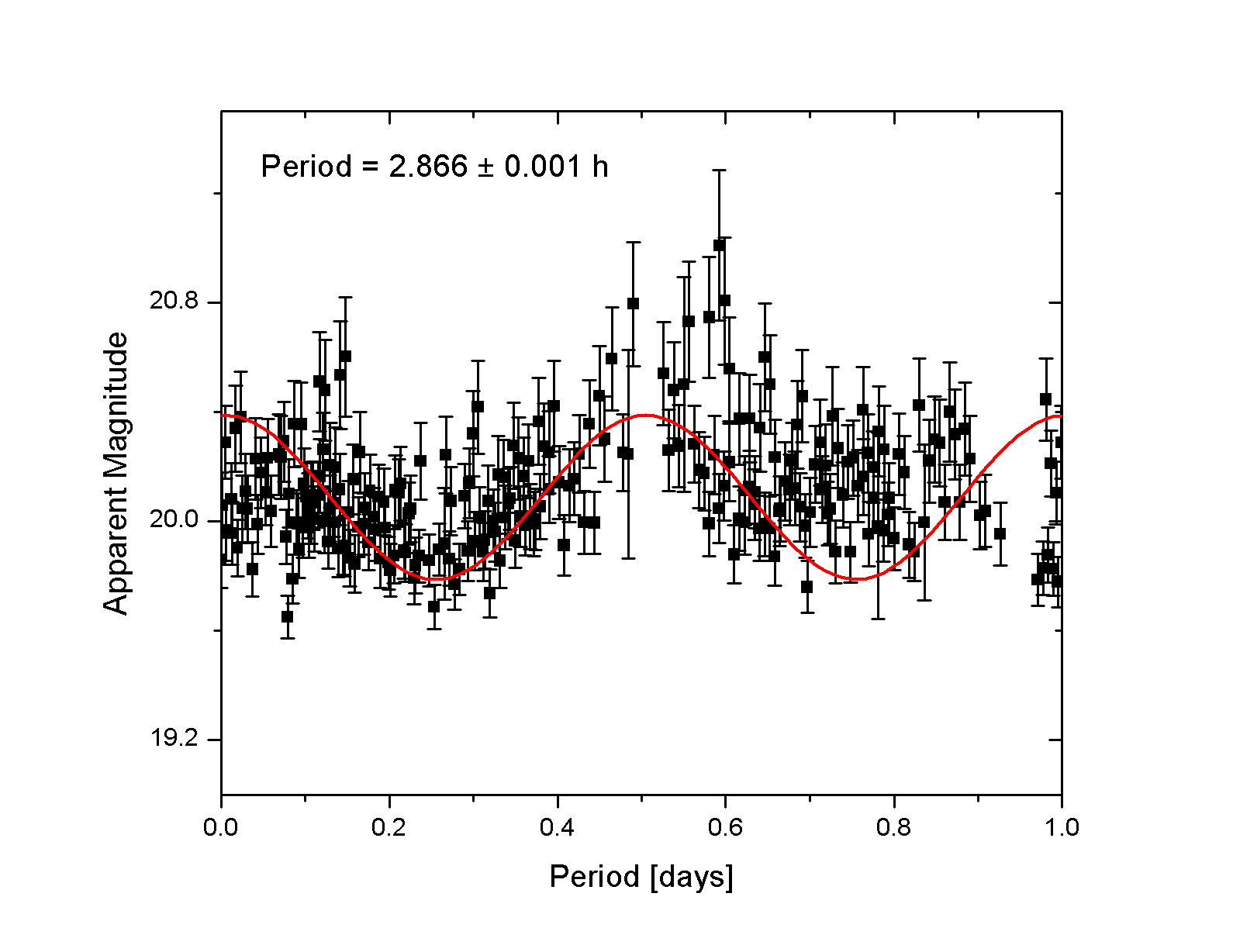}
\caption{}
\label{lightcurve2.2}
\end{subfigure}

	\caption{(259221)~2003~BA21 lightcurve. In Fig.~\ref{lightcurve2.1} is represented the most probable lightcurve with a period of 9.09 $\pm$ 0.02 h. In Fig.~\ref{lightcurve2} is represented the next probable lightcurve with a period of 2.866 $\pm$ 0.001 h}
		\label{lightcurve_259221}
\end{figure*}

The most probable period from our observations of (259221) 2003 BA21 is at 0.379 $\pm$ 0.001 days (9.09  $\pm$ 0.02 h). We also check the lightcurve for the next most probable period, that we found at 0.11942 $\pm$ 0.00002 days (2.866 $\pm$ 0.001 h).  The obtained lightcurves for those two periods are presented in Fig.~\ref{lightcurve_259221}.

	From the visual inspection we privilege the period at 9.09 $\pm$ 0.02 h, the period of 2.866 $\pm$ 0.001 h being too noisy. For a robust conclusion we need to investigate observational data that which stretch over several nights.

\subsection{Asteroids colors and reflectances}

For each asteroid we computed the colors $g-r$ and $g-i$ and reflectances (Table~\ref{colors_tab}) with the method presented in section~\ref{sec:color}.
\begin{table*}
\centering
\caption{Colors and the corresponding  spectral reflectances according to SDSS system.}
\label{colors_tab}
\begin{tabular}{ccccccccccc}\hline
Number  & Name      & g-r    & $\sigma_{g-r}$ & $R_{r}$    & $\sigma_{R_{r}}$  & g-i    & $\sigma_{g-i}$  & $R_{i}$    & $\sigma_{R_{i}}$\\ 
\hline
5141  & Tachibana & 0.2490 & 0.0568 & 1.0996 & 0.0227 & 0.3072 & 0.0513 & 1.1229 & 0.0205 \\
43032 & 1999 VR26 & 0.2539 & 0.0955 & 1.1016 & 0.0382 & 0.1880 & 0.1569 & 1.0752 & 0.0628 \\
85953 & 1999 FK21 & 0.0911 & 0.1454 & 1.0364 & 0.0581 & 0.0703 & 0.1600 & 1.0281 & 0.0640 \\
363599& 2004 FG11 & 0.1909 & 0.0826 & 1.0764 & 0.0331 & 0.1927 & 0.1024 & 1.0771 & 0.0409 \\
\hline
\end{tabular}
\end{table*}

\begin{figure*}
\centering
\includegraphics[width=0.9\linewidth]{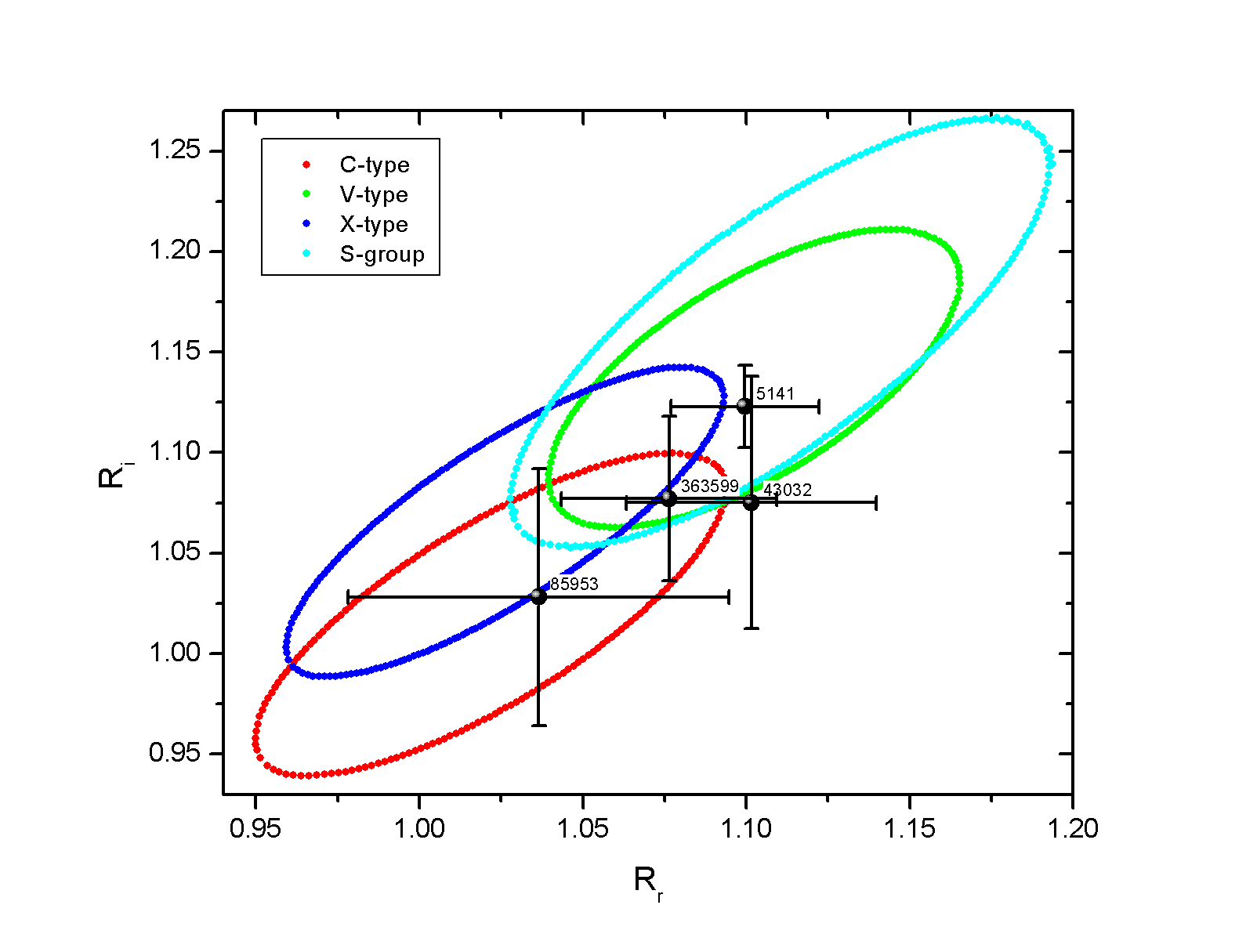}
\caption{ Asteroids taxonomic classes association. The S-group contain taxonomic classes A, L, S and Q-type. The taxonomic classes are based on SDSS-based Asteroid Taxonomy database.}
\label{taxon}
\end{figure*}

	In the next step we compared this values with SDSS-based asteroid Taxonomy\footnote{https://sbn.psi.edu/pds/resource/sdsstax.html} database.  We selected the most representative taxonomic classes \citep{2009Icar..202..160D}:   V, X-types and C-complex, S-groups (composed by A, L, S and Q-type). For each taxonomic class an ellipse borders the area of reflectances in the  ($R_r$, $R_i$) diagram. The values inferred for our objects are also displayed (Fig.~\ref{taxon})

	Asteroid (363599)~2004~FG11, classified as a V type,  may be associated with all the representative classes, but is most akin to a V-type, or S-group and less likely to be a C or X-type.

	Asteroid (85953)~1999~FK21, classified as S-type,  belongs most likely to a X or C taxonomic group (Fig.~\ref{taxon}).

	Finaly, asteroids (43032)~1999~VR26 and (5141) Tachibana  show an affinity to S-group or V-types (Fig.~\ref{taxon}).

\section{Conclusions}

In summary, from our data we infer :
\begin{itemize}
\item[--]
	For (363599)~2004~FG11, a rotation period of 7.021 $\pm$ 0.001 h, and a lightcurve compatible with the known binary system period of 22~h.
\item[--]
	For (259221)~2003~BA21, we obtain a preliminary rotational period of 9.1 $\pm$ 0.02 h, for a robust conclusion we need to investigate observational data which stretch over several nights. 
\item[--]
	Using colors obtained in the run and a comparison with SDSS database confirmation of a V-type class for (363599)~2004~FG11. (85953)~1999~FK is more akin to a C or X types than the S-type. The asteroids (43032)~1999~VR26 and (5141) Tachibana  are compatible with both V-type or S-group taxonomic type.
\end{itemize}

\acknowledgements

	This article contains data provided by observations run obtained using T1M at Pic du Midi Observatory, France.

	This research has made use of "Aladin sky atlas" developed at CDS, Strasbourg Observatory, France.

	This work was supported by Astronomical Institute of the Romanian Academy, Paris Observatory PICS--PASSO program, 
        Institute of Space Science and by a grant of the Romanian National Authority for Scientific Research
        UEFISCDI, project number PN--II--RU--TE--2014--4--2199.

        How to cite this article: Dumitru BA, Birlan M, Sonka A, Colas F, Nedelcu DA. Photometry of asteroids (5141), (43032), (85953), (259221), and (363599) observed at Pic du Midi Observatory. Astron. Nachr./AN. 2018;1–6. https://doi.org/10.1002/asna.201813463

\bibliographystyle{an}
\bibliography{an-demo}



\end{document}